\documentclass[10pt, conference, compsocconf]{IEEEtran}
\ifCLASSINFOpdf
\else
 \usepackage[dvips]{graphicx}
 \graphicspath{{./figures/}}
 \DeclareGraphicsExtensions{.eps}
\fi
 \usepackage{algpseudocode}
 \usepackage{listings}
 \usepackage{xcolor}
\usepackage{multirow}

\begin{document}
%
\title{Scaling Monte Carlo Tree Search on Intel Xeon Phi}


\author{\IEEEauthorblockN{S. Ali Mirsoleimani\IEEEauthorrefmark{1}\IEEEauthorrefmark{2},
Aske Plaat\IEEEauthorrefmark{1},
Jaap van den Herik\IEEEauthorrefmark{1} and
Jos Vermaseren\IEEEauthorrefmark{2}}
\IEEEauthorblockA{\IEEEauthorrefmark{1}Leiden Centre of Data Science, Leiden University\\
 Niels Bohrweg 1, 2333 CA Leiden, The Netherlands}
\IEEEauthorblockA{\IEEEauthorrefmark{2}Nikhef Theory Group, Nikhef\\
 Science Park 105, 1098 XG Amsterdam, The Netherlands}}


%


\maketitle

\begin{abstract}
Many algorithms have been parallelized successfully on the Intel Xeon Phi coprocessor, especially those with regular, balanced, and predictable data access patterns and instruction flows. Irregular and unbalanced algorithms are harder to parallelize efficiently. They are, for instance, present in artificial intelligence search algorithms such as Monte Carlo Tree Search (MCTS). In this paper we study the scaling behavior of MCTS, on a highly optimized real-world application, on real hardware. The Intel Xeon Phi allows shared memory scaling studies up to 61 cores and 244 hardware threads. We compare work-stealing (Cilk Plus and TBB) and work-sharing (FIFO scheduling) approaches. Interestingly, we find that a straightforward thread pool with a work-sharing FIFO queue shows the best performance. A crucial element for this high performance is the controlling of the grain size, an approach that we call \emph{Grain Size Controlled Parallel MCTS}. Our subsequent comparing with the Xeon CPUs shows an even more comprehensible distinction in performance between different threading libraries. We achieve, to the best of our knowledge, the fastest implementation of a parallel MCTS on the 61 core Intel Xeon Phi using a real application (47 relative to a sequential run).

\end{abstract}

\begin{IEEEkeywords}
Monte Carlo Tree Search, Intel Xeon Phi, Many-core, Scaling, Work-stealing, Work-sharing, Scheduling

\end{IEEEkeywords}

%
\IEEEpeerreviewmaketitle

\section{Introduction}
Modern computer architectures show an increasing number of
processing cores on a chip. There is a variety of many-core architectures. In
this paper we focus on the Intel Xeon Phi (Xeon Phi), which has currently 61 cores.
The Xeon Phi has
been designed for high degrees of parallelism with 244 hardware
threads and vectorization. Thus any application should generate and manage a large number of
tasks to use Xeon Phi efficiently. As long as all threads are executing balanced tasks, the tasks are easy to manage and each
application runs efficiently. If the 
tasks have imbalanced computations, a naive scheduling approach
may cause many threads to sit 
idle and thus decrease the system performance. However, many applications rely on unbalanced computations.

In this paper, we investigate how to parallelize irregular and
unbalanced tasks efficiently on Xeon Phi using MCTS. MCTS is an algorithm
for adaptive optimization that is used frequently in artificial
intelligence~\cite{Enzenberger2010a}\cite{Ruijl2014}\cite{Herik2013}\cite{Chaslot2008}\cite{Kocsis2006b}\cite{Mirsoleimani2015a}.
 
MCTS performs a search process based on a
large number of random samples in the search space. The nature of each sample in MCTS implies
that the algorithm is considered as a good target for 
parallelization. Much of the effort to parallelize MCTS has focused on using parallel threads
to do tree-traversal in parallel along separate paths in the search tree~\cite{Chaslot2008}\cite{Yoshizoe2011a}. Below we present a task parallelism approach with grain size control for MCTS.

We have chosen to use MCTS for the following three reasons. (1) Variants of Monte Carlo simulations
are a good benchmark for verifying the capabilities of
Xeon Phi architecture~\cite{Shuo-li2013a}. (2) The scalability of parallel MCTS is a challenging task for Xeon Phi~\cite{Mirsoleimani2014b}. (3) MCTS searches the tree
asymmetrically~\cite{Kuipers2013}, making it an interesting 
application for investigating the parallelization of tasks with 
irregular and unbalanced computations~\cite{Mirsoleimani2014b}. 

Three main contributions of this paper are as follows.
\begin{enumerate} 
\item A first detailed analysis of the performance of three
 widely-used threading libraries on a highly optimized program with high levels of
 irregular and unbalanced tasks on Xeon Phi is provided.
\item A straightforward First In First
 Out (FIFO) scheduling policy is shown to be equal or even to outperform the more elaborate threading
 libraries Cilk Plus and TBB for running high levels of parallelism for high numbers of cores. This is surprising since Cilk Plus was designed to achieve high
 efficiency for precisely these types of applications. 
\item A new parallel MCTS with grain size control is proposed. It achieves, to the best of our knowledge, the fastest
 implementation of a parallel MCTS on the 61-core Xeon Phi 7120P
 (using a real application) with 47 times speedup compared to
 sequential execution on Xeon Phi itself (which translates to 5.6
 times speedup compared to the sequential version on the regular host
 CPU, Xeon E5-2596).
\end{enumerate}

The rest of this paper is organized as follows. In section \ref{sec:back} the
required background information is briefly discussed. Section \ref{sec:gscpm} describes the grain size control parallel MCTS. Section \ref{sec:setup} provides the experimental setup, and section \ref{sec:results} gives the experimental results. Section \ref{sec:discuss} presents the analysis of results. Section \ref{sec:related} discusses related work. Finally, in Section \ref{sec:conclusion} we
conclude the paper. 

\section{Background}
\label{sec:back}
Below we provide some background of four topics. parallel programming models (Section \ref{sec:models} ), MCTS (Section \ref{sec:mcts} ), the game of Hex (Section \ref{sec:hex} ), and the architecture of the Xeon Phi (Section \ref{sec:phi}).
\subsection{Parallel Programming Models}
\label{sec:models}

Some parallel programming models provide programmers with 
thread pools, relieving them of the need to manage their
parallel tasks explicitly~\cite{cilk1998}\cite{cilkplus2015}\cite{openmp2015}.
Creating threads each time that a program needs them can be
undesirable. To prevent overhead, the program has to manage the
lifetime of the thread objects and to determine the number of threads
appropriate to the problem and to the current hardware. The ideal
scenario would be that the program could just (1) divide the code into
the smallest logical pieces that can be executed concurrently (called tasks),
(2) pass them over to the compiler and library, in order to
parallelize them. This approach uses the fact that the majority of
threading libraries do not destroy the threads once created, so they
can be resumed much more quickly in subsequent use. This 
is known as creating a \emph{thread pool}. 

A thread pool is a group of shared threads~\cite{Nichols1996}. Tasks that can be executed
concurrently are submitted to the pool, and are added to a queue of
pending work. Each task is then taken from the queue by one of the
worker threads, that execute the task before looping back to take
another task from the queue. The user specifies the number of worker
threads.

Thread pools use
either a a work-stealing or a work-sharing scheduling method to balance
the work load. Examples of parallel programming models with work-stealing
scheduling are TBB~\cite{tbb2015} and Cilk
Plus~\cite{cilkplus2015}. The work-sharing method
is used in the OpenMP programming model~\cite{openmp2015}. 

Below we discuss two threading libraries: Cilk
Plus and TBB. 

\subsubsection{ Cilk Plus}
Cilk Plus is an extension to C and C++ designed to offer a quick and
easy way to harness the power of both multicore and vector
processing. Cilk Plus is based on MIT's research on Cilk~\cite{Blumofe1995}. Cilk Plus provides a simple yet 
powerful model for parallel programming, while runtime and template
libraries offer a well-tuned environment for building parallel
applications~\cite{6478757}. 

Function calls can be tagged with the first keyword \emph{cilk\_spawn} which
indicates that the function can be executed concurrently. The calling function uses the second keyword \emph{cilk\_sync} 
to wait for the completion of all the functions it spawned. The third keyword is \emph{cilk\_for} which converts a simple for loop into a parallel for loop. The tasks are executed by the runtime system
within a provably efficient work-stealing framework. Cilk Plus uses a double ended queue per thread to keep track of the tasks to execute and uses it as a stack
during regular operations conserving a sequential semantic. When a
thread runs out of tasks, it steals the deepest half of the stack of
another (randomly selected) thread~\cite{cilk1998}~\cite{6478757}. In Cilk Plus, thieves steal \emph{continuations}.

\subsubsection{Threading Building Blocks}
 Threading Building Blocks (TBB) is a C++ template library
developed by Intel for writing software programs that take advantage
of a multicore processor~\cite{tbb-reinders2007-book}. 
TBB implements work-stealing to balance a parallel workload across
available processing cores in order to increase core utilization and
therefore scaling. The TBB work-stealing model is similar to the work
stealing model applied in Cilk, although in TBB, thieves steal \emph{children}~\cite{tbb-reinders2007-book}. 

\subsection{Monte Carlo Tree Search}
\label{sec:mcts}
MCTS is a tree search method that has been
successfully applied in games such as Go, Hex and other applications
with a large state space~\cite{Herik2013}\cite{Coulom2006}\cite{Arneson2010}. It
works by selectively building a tree, expanding only branches it deems 
worthwhile to explore. MCTS consists of four steps~\cite{chaslot2008progressive}. (1) In the selection step, a leaf (or a not fully expanded node) is selected according to some criterion. (2) In the expansion
step, a random unexplored child of the selected node is added to the
tree. (3) In the simulation step (also called playout), the rest of the path to a final node is
completed using random child selection. At the end a score $\Delta$ is
obtained that signifies the score of the chosen path through the state
space. (4) In the backprogagation step (also called backup step), this
value is propagated back through the tree, which affects the average
score (win rate) of a node. The tree is built iteratively by repeating
the four steps. In the games of Hex and Go, each node represents a
player move and in the expansion phase the game is played out, in
basic implementations, by random moves. Figure \ref{alg:mcts} shows the general MCTS algorithm. In many MCTS implementations the Upper Confidence Bounds for Trees (UCT) is chosen as the selection criterion~\cite{Kocsis2006b}\cite{Browne2012} for the trade-off between
exploitation and exploration that is one of the hallmarks of the algorithm. 

\begin{figure}
\begin{algorithmic}
\Function{UCTSearch}{r,m}
\State i$\gets$ 1
\For{i$\leq$ m}
 \State n$\gets$ select(r)
 \State n$\gets$ expand(n)
 \State $\Delta \gets$playout(n)
 \State backup(n,$\Delta$)
 \EndFor
\State \Return
\EndFunction
\end{algorithmic}
\caption{The general MCTS algorithm}
\label{alg:mcts}
\end{figure}

The UCT algorithm addresses the problem of balancing exploitation and exploration in the selection phase of the MCTS
algorithm~\cite{Kocsis2006b}. A child node $j$ is selected to maximize: 
\begin{equation}
UCT(j)=\overline{X}_{j}+C_{p}\sqrt{\frac{\ln(n)}{n_{j}}}
\end{equation}
where $\overline{X}_{j}=\frac{w_{j}}{n_{j}} $, $w_{j}$ is the number
of wins in child $j$, $n_{j}$ is the number of times child $j$ has
been visited, $n$ is the number of times the parent node has been
visited, and $C_{p}\geq0$ is a constant. The first term in the UCT
equation is for exploitation of known parts of the tree and the second one is for
exploration of unknown parts~\cite{Browne2012}. The level of exploration of the UCT
algorithm can be adjusted by the $C_{p}$ constant. 

\subsection{Hex Game}
\label{sec:hex}
Previous scalability studies used artificial trees~\cite{Yoshizoe2011a}, simulated
hardware~\cite{Segal:2010:SPU:1950322.1950326}, or a large and complex
program~\cite{Mirsoleimani2014a}. All of these approaches suffered
from some drawback for performance profiling and analysis. For this reason, we developed from
scratch a program with the goal of generating realistic trees 
while being sufficiently transparent to allow good performance analysis, using 
the game of Hex. 

Hex is a game with a board of hexagonal cells
~\cite{Arneson2010}. Each player is represented by a color (White or
Black). Players take turns by alternatively placing a stone of their color on a cell
of the board. The goal for each player is to create a connected
chain of stones between the opposing sides of the board
marked by their colors. The first player to complete this path
wins the game.

In our implementation of the Hex game, a disjoint-set data structure
is used to determine the connected stones. Using this data structure
we have an efficient representation of the board position to find the player who won the game~\cite{Galil:1991:DSA:116873.116878}. 

\subsection{Architecture of the Intel Xeon Phi}
\label{sec:phi}
\begin{figure}[!t]
\centering
\includegraphics[scale=0.35]{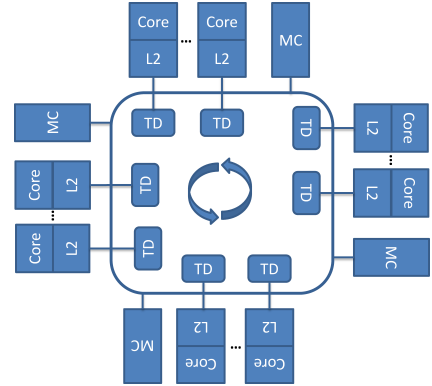}
\caption{Abstract of Intel Xeon Phi microarchitecture}
\label{fig:phi}
\end{figure}

We will now provide a brief overview of the Xeon Phi co-processor
architecture (see Figure~\ref{fig:phi}). A Xeon Phi co-processor board
consists of up to 61 cores based on the Intel 64-bit ISA. Each of these cores contains \emph{vector processing units} (VPU) to 
execute 512 bits of 8 double-precision floating point elements or 16 single-precision floats or 32-bit integers at the same time, 4-way SMT, and dedicated L1 and fully coherent L2
 caches~\cite{Rahman2013}. The \emph{tag directories} (TD) are
 used to look up cache data distributed among the cores. The
 connection between cores and other functional units such as
 \emph{memory controllers} (MC) is through a bidirectional
 \emph{ring interconnect}. There are 8 distributed memory
 controllers as interface between the ring burst and main memory
 which is up to 16 GB. 

\section{Grain Size Controlled Parallel MCTS}
\label{sec:gscpm}
\begin{figure}
\centering
\includegraphics[scale=0.7]{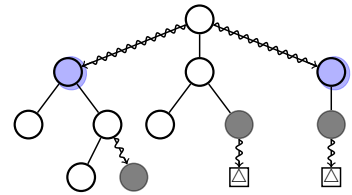}
\caption{Tree parallelism with local lock. The curly arrows represent threads. The grey nodes are locked ones. The dark nodes are newly added to the tree.}
\label{fig:PMCTS-1}
\end{figure}

The literature describes three main methods for parallelizing MCTS on shared memory machines: leaf
parallelism, root parallelism, and tree parallelism~\cite{Chaslot2008}\cite{Mirsoleimani2015a}\cite{Mirsoleimani2014a}. 
In this paper an algorithm based on tree parallelism is proposed.
In tree parallelism one MCTS tree is shared among
several threads that are performing simultaneous tree-traversal starting from the root~\cite{Chaslot2008}. Figure \ref{fig:PMCTS-1} shows the tree parallelism
algorithm with local locks.

It is often hard to find sufficient parallelism in an application when there is a large number of cores available as in Xeon Phi. To find enough parallelism, we will adapt MCTS to use logical parallelism, called tasks (see Figure \ref{alg:pmcts}).
In the MCTS algorithm, iterations can be divided into chunks to be executed serially. A \emph{chunk} is a sequential collection of one or more iterations. The maximum size of a chunk is called \emph{grain size}. Controlling the number of tasks (\emph{nTasks}) allows to control the grain size (\emph{m}) of our algorithm. The runtime can allocate the tasks to threads for execution. With this technique we can create more tasks than threads i.e., fine-grained parallelism. By reducing the grain size we expose more parallelism to the compiler and threading library~\cite{Reinders2014}. Finding the right balance is the key to achieve a good scaling. The grain size should not be too small because then spawn overhead reduces performance. It also should not be too large because that reduces parallelism and load balance (see Table~\ref{tab:gsize}).

\begin{table}
\centering
\caption{The conceptual effect of grain size.}
\label{tab:gsize}
\begin{tabular}{c|c}
\hline
\begin{tabular}[c]{@{}c@{}}Large grain size\\ (nTasks $\ll$ nCores)\end{tabular}       & \begin{tabular}[c]{@{}c@{}}Speedup bounded by tasks\\  (not enough parallelism)\end{tabular}   \\ \hline
Right grain size                                                                                      & Good speedup                                                                                   \\ \hline
\begin{tabular}[c]{@{}c@{}}Small grain size\\ (nTasks $\gg$ nCores)\end{tabular} & \begin{tabular}[c]{@{}c@{}}Spawn and scheduling overhead \\ (reduces performance)\end{tabular} \\ \hline
\end{tabular}
\end{table}

In order to change the grain size, a wrapper function needs to be created. In this function, the maximum number of playouts is divided by the number of desired tasks. Each individual task has $nPlayouts/nTasks$ iterations. Then, the \emph{UCTSearch} is run for the specified number of iterations as a separate task. The grain size could be as small as one iteration.
We call this approach Grain Size Controlled Parallel MCTS (GSCPM).The GSCPM pseudo-code is shown in Figure \ref{alg:pmcts}.

\begin{figure}
\begin{algorithmic}
\Function{GSCPM}{$s$,$nPlayouts$}
\State $m \gets nPlayouts/nTasks$
\State $r\gets$ create a shared root node with state s
\State $t \gets 1$
\For{$t\leq nTasks$}
	\State execute UCTSearch($r$,$m$) as task $t$
\EndFor
\State wait for all tasks to be finished
\State \Return best child of $r$
\EndFunction
\end{algorithmic}
\caption{The pseudo-code of GSCPM algorithm.}
\label{alg:pmcts}
\end{figure}

\begin{figure}
\centering
\includegraphics[width=8cm]{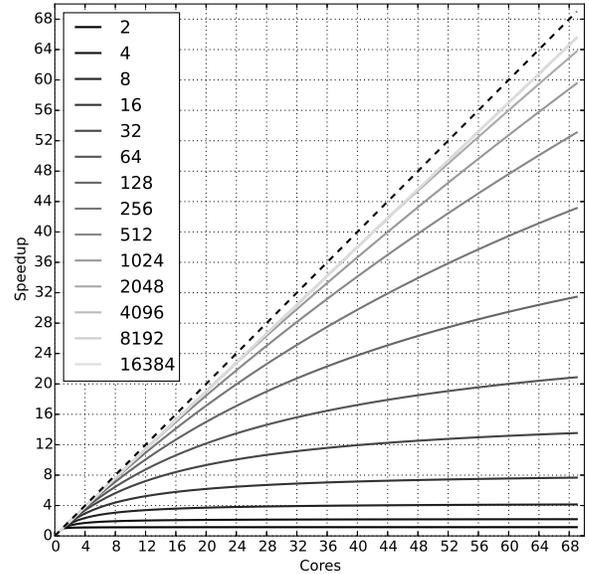}
\caption{The scalability profile produced by Cilkview for the GSCPM algorithm from Figure~\ref{alg:pmcts}. The number of tasks are shown. Higher is more fine-grained.}
\label{fig:cilkview}
\end{figure}

It is important to study how our method scales as the number of processing cores increases. Figure \ref{fig:cilkview} shows the scalability profile produced by Cilkview~\cite{He2010} that results from a single instrumented serial run of the GSCPM algorithm for different numbers of tasks. The curves show the amount of available parallelism in our algorithm; they are lower bounds indicating an estimation of the potential program speedup with the given grain size. As can be seen, fine-grained parallelism (many tasks) is needed for MCTS to achieve good intrinsic parallelism. Using 16384 tasks shows near perfect speedup on 61 cores. The actual performance of a parallel application is determined not only by its intrinsic parallelism, but also by the performance of the runtime scheduler. Therefore, it is important to find an efficient scheduler.

\subsection{Implementation Details}
\label{sec:impl}
Parallelization of the computational loop in the \emph{UCTSearch}
function (see Figure \ref{alg:mcts}) across processing units
(e.g., cores and threads) is rather simple. We assume the absence of any
logical dependencies between iterations.\footnote{This assumption may affect the quality of the search, but this phenomenon, called search overhead, is out of scope of this paper~\cite{Mirsoleimani2015a}} In our implementation of tree
parallelism, a lock is used in the expansion phase of the MCTS
algorithm in order to avoid the loss of any information and corruption
of the tree data structure~\cite{Enzenberger2010a}. To allocate all
children of a given node, a pre-allocated vector of children is
used. When a thread tries to append a new child to a node it
increments an atomic integer variable as the index to the next possible
child in the vector of children. The values of $w_{j}$ and $n_{j}$ are
also defined to be atomic integers (see Section \ref{sec:mcts}). 

The main approach to parallelize the algorithm is to create a number of parallel
tasks equal to \emph{nTasks}. By increasing the value of
\emph{nTasks}, each of the tasks contains a lower number of
iterations. Figure \ref{lst:code} shows how the algorithm is parallelized with different threading libraries. 
\begin{figure}
\begin{lstlisting}
//C++11
std::vector<std::thread> threads;
for (int t = 0; t < nTasks; t++) {
 threads.push_back(std::thread(UCTSearch(r,m)));
}
//Wait for all tasks to be finished

//TBB
tbb::task_group g;
for (int t = 0; t < nTasks; t++) {
	g.run(UCTSearch(r,m));
}
g.wait();

//Cilk Plus 1
for (int t = 0; t < nTasks; t++) {
	cilk_spawn UCTSearch(r,m);
}
cilk_sync;

//Cilk Plus 2
cilk_for (int t = 0; t < nTasks; t++) {
	UCTSearch(r,m);
}
//Wait for all tasks to be finished

//Thread pool with FIFO scheduling
for (int t = 0; t < nTasks; t++) {
	TPFIFO.schedule(UCTSearch(r,m));
}
TPFIFO.wait();
\end{lstlisting}
\caption{Task parallelism for GSCPM, based on different threading libraries.}
\label{lst:code}
\end{figure}

Cilk Plus and TBB are designed to efficiently balance the load for fork-join parallelism automatically, using a technique called work-stealing. In a basic work-stealing scheduler, each thread is called a worker. Each worker maintains its own double-ended queue (deque) of tasks. Cilk Plus and TBB differ in their concept of what is a stealable task. 
For each spawned \emph{UCTSearch}, there are two conceptual tasks: (1) the continuation of executing the loop around \emph{UCTSearch}, (2) a \emph{child} task \emph{UCTSearch}. This task called \emph{continuation}. A key difference between Cilk Plus and TBB is that in Cilk Plus, thieves steal \emph{continuations}. In TBB, thieves steal \emph{children}. In TPFIFO the tasks are put in a queue. It implements work-sharing; but the order that the tasks are executed is similar to \emph{child stealing}. The first task that enters the queue is the first task that gets executed.

In our thread pool implementation\footnote{We use the open source library based on Boost C++ libraries which is available at http://sourceforge.net/projects/threadpool/} (called TPFIFO) the task
functions are executed asynchronously. A task is submitted to a FIFO
work queue and will be executed as soon as one of the pool's
threads is idle. \emph{Schedule} returns immediately and there are no
guarantees about when the tasks are executed or how long the
processing will take. Therefore, the program waits for all the tasks
to be finished. 

Efficiency of Random Number Generation (RNG) is a crucial performance
aspect of any Monte Carlo simulation~\cite{Shuo-li2013}. In our implementation, the highly optimized Intel MKL is used to
generate a separate RNG stream for each task. One MKL RNG interface API
call can deliver an arbitrary number of random numbers. In our program,
a maximum of 64 K random numbers are delivered in one call~\cite{Wang2014}. A thread generates the required number of random numbers for each task.

\section{Experimental Setup}
\label{sec:setup}
The goal of this paper is to study the scalability of
irregular unbalanced task parallelism on the Xeon Phi. We do so using a
specially written, highly optimized, Hex playing program, in order to generate realistic
real-world search spaces.\footnote{Source code is available at https://github.com/mirsoleimani/paralleluct/}

\subsection{Test Infrastructure}
The performance evaluation of GSCPM was performed on a dual socket Intel machine with 2 Intel {\em Xeon\/} E5-2596v2 CPUs running at 2.40GHz. Each CPU has 12 cores, 24 hyperthreads and 30 MB L3 cache. Each physical core has 256KB L2 cache. The peak TurboBoost frequency is 3.2
GHz. The machine has 192GB physical memory. The machine is equipped with an Intel {\em Xeon Phi\/} 7120P 1.238GHz which has 61 cores and 244
hardware threads. Each core has 512KB L2 cache. The Xeon phi has
16GB GDDR5 memory on board with an aggregate theoretical bandwidth of
352 GB/s.

The Intel Composer XE 2013 SP1 compiler was
used to compile for both Intel Xeon CPU and Intel Xeon Phi. Four different threading libraries were used for
evaluation: C++11, Boost C++ libraries 1.41, Intel
Cilk Plus, and Intel TBB 4.2. We compiled the code using the Intel C++ Compiler with a
-$O3$ key. 

In order to generate statistically significant results in a reasonable
amount of time, 1,048,576 playouts are executed to choose a move. The
board size is 11x11. The UCT constant $C_p$ is set at 1.0 in all of our experiments. To
calculate the playout speedup the average of time
over 10 games is measured for doing the first move of the game when
the board is empty. The results are within less than 3\% standard deviation.

\section{Experimental Results}
\label{sec:results}
Table \ref{tab:seq_time} shows the sequential time to execute the
specified number of playouts. The sequential time on
the Xeon Phi is
almost 8 times slower than the Xeon CPU. This is because each core on
the Xeon Phi is slower than each one on the Xeon CPU. (The Xeon Phi has
in-order execution, the CPU has out-of-order execution, hiding the
latency of many cache misses.)

\begin{table}[!t]
\renewcommand{\arraystretch}{1.3}
\caption{Sequential version. Time in seconds.}
\label{tab:seq_time}
\centering
\begin{tabular}{l|c|c}
\hline
Processor & Board Size & Sequential Time (s) \\ \hline
Xeon CPU & 11x11 & $21.47\pm0.07$ \\ \hline
Xeon Phi & 11x11 & $185.37\pm0.53$ \\ \hline
\end{tabular}
\end{table}

The time of execution in the first game is rather longer on
Xeon Phi; therefore the overhead costs for thread creation may include
a significant contribution to the parallel region execution
time. This is a known feature of the Xeon Phi, called the warm-up
phase~\cite{Reinders2014}. Therefore, the first game is not included
in the results to remove that overhead. The majority of threading library implementations do not destroy the
threads created for the first time~\cite{Reinders2014}. 

\begin{figure}[!t]
\centering
\includegraphics[width=8cm]{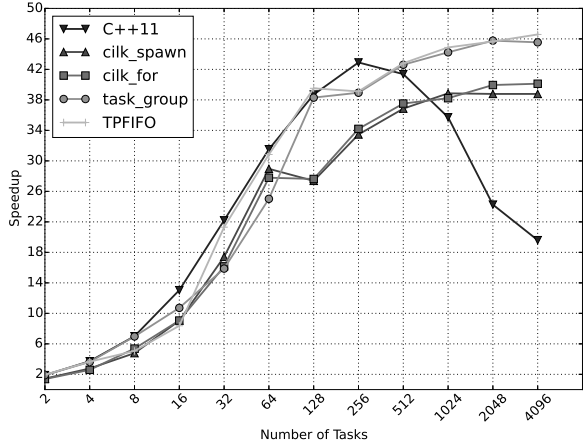}
\caption{Speedup on Intel Xeon Phi. Higher is better. Left: coarse-grained parallelism. Right: fine-grained parallelism.}
\label{fig:phi_time}
\end{figure}

The graph in Figure \ref{fig:phi_time} shows the speedup for
different threading libraries on a Xeon Phi, as a function of the number
of tasks. We recall that going to the right of the graph, finer grain parallelism is observed. 

Creating threads in C++11 equal to the number of tasks is the simplest approach. The best speedup for C++11 on Xeon Phi is achieved for 256 threads/tasks. It is better than \emph{cilk\_spawn} and \emph{cilk\_for} from Cilk Plus and TBB (\emph{task\_group}) with the same number of tasks. However, the limitation of this approach is that creating larger numbers of threads has large overhead. The reduction in speedup for C++11 is shown in Figure~\ref{fig:phi_time}.

For \emph{cilk\_for} on the Xeon Phi, the
best execution time is achieved for fine-grained tasks, when the number of tasks is greater than
2048. The performance of TBB (\emph{task\_group}) and TPFIFO are
quite close on the Xeon Phi. TPFIFO scales for up to 4096 tasks and  TBB (\emph{task\_group}) scales for up to 2048 tasks. 
The reason for the similarity between TBB and TPFIFO on Xeon Phi is explained in \ref{sec:impl}. 

\section{Discussion and Analysis}
\label{sec:discuss}
\begin{figure}[!t]
\centering
\includegraphics[width=8cm]{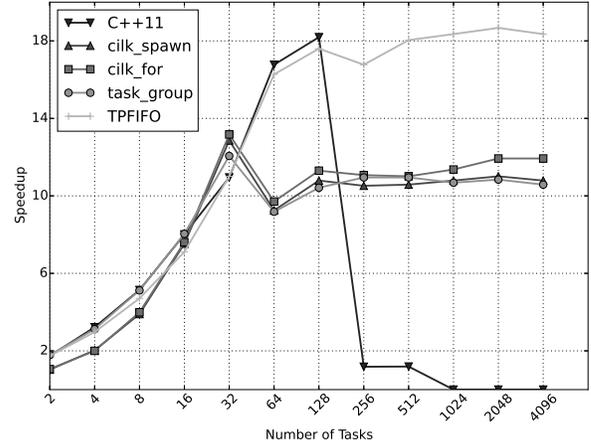}
\caption{Speedup on Intel Xeon CPU. Higher is better. Left: coarse-grained parallelism. Right: fine-grained parallelism.}
\label{fig:cpu_time}
\end{figure}

In analyzing our results on Xeon Phi, we also measured the performance of GSCPM on Xeon CPU. 
Figure \ref{fig:cpu_time} depicts the observed time for each threading
library as a function of the number of processed tasks, as measured on
Xeon CPU. We see three interesting facts. (1) Each method has different behaviors, which depend on its
implementation. It is shown that on the Xeon CPU, by doubling the numbers of
tasks the running time becomes almost half for up to 32 threads for
C++11, Cilk Plus (\emph{cilk\_spawn} and \emph{cilk\_for}), and TPFIFO. C++11 and TBB (\emph{task\_group}) achieve very
close performance for up to 32 tasks. (2) For 64 and 128 tasks, the speedup for C++11 is better than for Cilk Plus and TBB.
Cilk Plus speedup is less than the other 
methods up to 16 threads. (3) The best time for \emph{cilk\_for} on Xeon CPU is
observed for coarse-grained tasks, when the numbers of tasks are equal to 32. It shows the optimal task grain size for Cilk Plus.
The measured time for Cilk Plus comes very close to TBB (\emph{task\_group})  on Xeon CPU, while it never reaches to TBB (\emph{task\_group}) performance on Xeon Phi after 64 tasks.

\begin{figure}[!t]
\centering
\includegraphics[width=8cm]{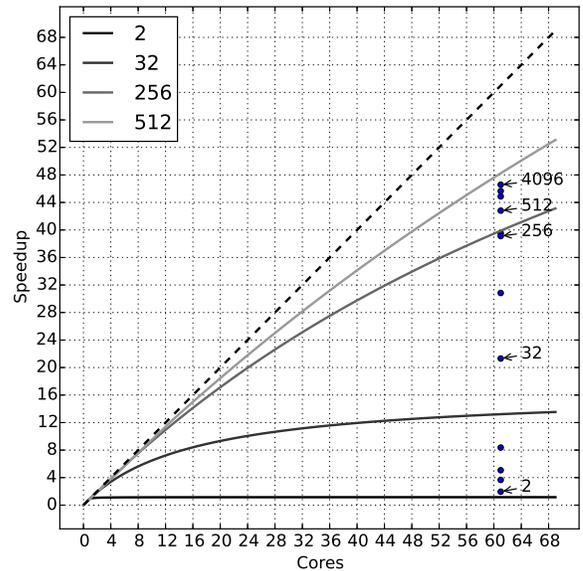}
\caption{Comparing Cilkview analysis with TPFIFO speedup on Xeon Phi. The dots show the number of tasks used for TPFIFO. The lines shows the number of tasks used for Cilkview. }
\label{fig:phi_cv_tp_cmp}
\end{figure}

Figure \ref{fig:phi_cv_tp_cmp} shows a mapping from TPFIFO speedup to the Cilkview graph for 61 cores. We remark that the results of Figure~\ref{fig:phi_time} correspond nicely to the Cilkview results for up to 256 tasks. After that the speedup continues to improve but not as expected by Cilkview due to overheads. 

\section{Related Work}
\label{sec:related}
 
Saule et al.~\cite{Saule2012} compared the scalability of Cilk Plus, TBB, and OpenMP for a parallel graph coloring algorithm. They also studied the performance of aforementioned programming models for a micro-benchmark with irregular computations. The micro-benchmark is a parallel \emph{for} loop especially designed to be less memory intensive than graph coloring. The maximum speedup for this micro-benchmark on Xeon Phi was 47 and is obtained using 121 threads. 

Authors in~\cite{Tousimojarad2015} used a thread pool with work-stealing and compared it to OpenMP, Cilk Plus, and TBB. They used Fibonacci as an example of unbalanced tasks. 
In contrast to our results, their approach shows no improvement over other methods for unbalanced tasks before using 2048 tasks with work-stealing.


Baudi\v{s} et al.~\cite{citeulike:10600156} reported the performance of a lock free tree parallelism for up to 22 threads. They used a different speedup measure. The strength speedup is good up to 16 cores but the improvement drops after 22 cores. 

Yoshizoe et al.~\cite{Yoshizoe2011a} study the scalability of MCTS algorithm on distributed systems. They have used artificial game trees as the benchmark. Their closest settings to our study are 0.1 ms playout time and branching factor of 150 with 72 distributed processors. They showed a maximum of 7.49 times speedup for distributed UCT on 72 CPUs. They have proposed depth first UCT and reached 46.1  times speedup for the same number of processors.

\section{Conclusion}
\label{sec:conclusion}

In this paper we have presented a scaling study of irregular and unbalanced task parallelism on real hardware, using a full-scale, highly optimized, artificial intelligence program, a program to play the game of Hex, that was specifically designed from scratch for this study. 


We have studied a range of scheduling methods, ranging from a simple FIFO work queue, to state of the art work-sharing and work-stealing libraries. Cilk Plus and TBB are specifically designed for irregular and unbalanced (divide and conquer) parallelism. Despite the high level of optimization of our sequential code-base, we achieve good scaling behavior, a speedup of 47 on the 61 cores of the Xeon Phi. Surprisingly, this performance is achieved using one of the simplest scheduling mechanisms, a FIFO thread pool. 

In our analysis we found the notion of grain size to be of central importance. Traditional parallelizations of MCTS (tree parallelism, root parallelism) use a one-to-one mapping of the logical tasks to the hardware threads (see, e.g.,~\cite{Chaslot2008}\cite{Mirsoleimani2015a}\cite{Mirsoleimani2014b}). In this paper we introduced Grain Size Controlled MCTS, where the logical task is divided into more, smaller, tasks, that a straightforward FIFO work-sharing scheduler can then schedule efficiently, without the overhead of work-stealing approaches. 

A crucial insight of our work is to view the search job not as coarse-grained monolithic tasks, but as a logical task that consists of many individual searches which can be grouped into finer grained tasks. Note that our one-task-per core speedup reaches 31 on 61 cores and our many-tasks speedup reaches 47 on 61 cores (see Figure 7). We have not seen better performance on Xeon Phi.

For future work, here we remark that the view of a parallel search algorithm as consisting of loosely connected individual searches is reminiscent of Transposition Driven Scheduling \cite{Yoshizoe2011a}\cite{Romein99} and may find application in other best-first algorithms, e.g., in \cite{Huang2015}\cite{Plaat1994}. Therefore, we envision a wider applicability of Grain Size Controlled to other best-first searches. Furthermore, the high scaling performance is promising. Our approach with a specially designed clean program allows easy experimentation, and we are working on extending the thread pool scheduler on the Xeon Phi.

\section*{Acknowledgment}
The authors would like to thank Jan Just Keijser for his helpful
support with the Intel Xeon Phi co-processor. 
This work is supported in part by the ERC Advanced Grant no. 320651, “HEPGAME.”



\bibliographystyle{IEEEtran}
\bibliography{IEEEabrv,Bib-icpads2015}

%
%
%

\end{document}